\documentclass[pdflatex,sn-mathphys-num]{sn-jnl}

\usepackage{graphicx}%
\usepackage{multirow}%
\usepackage{amsmath,amssymb,amsfonts}%
\usepackage{amsthm}%
\usepackage{mathrsfs}%
\usepackage[title]{appendix}%
\usepackage{xcolor}%
\usepackage{textcomp}%
\usepackage{manyfoot}%
\usepackage{booktabs}%
\usepackage{algorithm}%
\usepackage{algorithmicx}%
\usepackage{algpseudocode}%
\usepackage{listings}%

\theoremstyle{thmstyleone}%
\theoremstyle{thmstyletwo}%

\theoremstyle{thmstylethree}%
\newtheorem{definition}{Definition}%

\begin{document}

\title[Article Title]{Occupation Life Cycle}


\author[1]{\fnm{Lan} \sur{Chen}}

\author[1]{\fnm{Yufei} \sur{Ji}}

\author[1]{\fnm{Xichen} \sur{Yao}}

\author*[1]{\fnm{Hengshu} \sur{Zhu}}\email{zhuhengshu@kanzhun.com}
\affil[1]{\orgdiv{Career Science Lab}, \orgname{BOSS Zhipin}}


\abstract{This paper explores the evolution of occupations within the context of industry and technology life cycles, highlighting the critical yet underexplored intersection between occupational trends and broader economic dynamics. Introducing the Occupation Life Cycle (OLC) model, we delineate five stages—growth, peak, fluctuation, maturity, and decline—to systematically explore the trajectory of occupations. Utilizing job posting data from one of China's largest recruitment platforms as a novel proxy, our study meticulously tracks the fluctuations and emerging trends in the labor market from 2018 to 2023. Through a detailed examination of representative roles, such as short video operators and data analysts, alongside emerging occupations within the artificial intelligence (AI) sector, our findings allocate occupations to specific life cycle stages, revealing insightful patterns of occupational development and decline. Our findings offer a unique perspective on the interplay between occupational evolution and economic factors, with a particular focus on the rapidly changing Chinese labor market. This study not only contributes to the theoretical understanding of OLC but also provides practical insights for policymakers, educators, and industry leaders facing the challenges of workforce planning and development in the face of technological advancement and market shifts.
}




\maketitle

\section{Introduction}\label{sec1}

Scholars have extensively explored the evolution of products and industries~\cite{utterback1975dynamic,klepper1997industry}, yet the evolution of occupations has received comparatively less attention. This oversight is notable given the profound impact occupational shifts have on economic landscapes and labor markets. Analyzing the evolution of occupations is crucial for several reasons. It provides insights into the dynamics of the labor market, informs policy-making to address employment challenges, and helps predict future demands for skills and training. Understanding occupational changes offers a comprehensive view of how technological advancements and industry transformations influence the workforce and employment opportunities.

In this study, we examine the influences of the occupation life cycle (OLC) and its interrelation with other pivotal cycles, namely the industry life cycle (ILC)~\cite{utterback1975dynamic,klepper1997industry} and the technology life cycle (TLC)~\cite{linden2003understanding,dedehayir2016hype}. Our analysis seeks to identify the fundamental factors affecting occupational development from both the demand and supply sides. By examining characteristic recruitment behaviors, we define five distinct stages of the OLC: growth, peak, fluctuation, maturity, and decline. To elucidate this concept, we construct a visualization that captures the trend of the OLC, attempting to pinpoint the current stage of each occupation.

Recognizing that labor market patterns and occupational trends vary across countries, we select China as a case study. This choice is motivated by China's significant global economic presence and its dynamic labor market, which exemplifies the complexities and rapid changes characteristic of large economies. We gathered data from one of China's largest recruiting platforms, tracking the number of job postings for each occupation from 2018 to 2023. Given the scarcity of long-term data, our focus primarily rests on analyzing occupations that have experienced substantial fluctuations during this period or those that are newly emerging.

This investigation into China's labor market through the lens of the OLC offers a novel perspective on occupational evolution. By dissecting the intricate relationship between occupational trends and the broader economic and technological contexts, our study contributes to a deeper understanding of workforce dynamics in the face of rapid global changes.

\section{Theoretical Background}\label{sec2}

In this section, we review the literature on life cycle phenomena regarding various economic behaviors, in particular, the relationship between occupation life cycle patterns and other life cycle patterns. We argue an integrative and interdisciplinary perspective is required to construct the concept of OLC.
Furthermore, to understand the driving factors behind OLC, we also provide an economic analysis of the occupational development from both sides of the labor market.

\subsection{OLC and other life cycle phenomena}

The life cycle patterns are stylized phenomena that are often observed in various types of economic systems.
The concept of OLC emerges as a multifaceted phenomenon intricately linked to the cyclical dynamics observed across different levels of economic and human behaviors.
This perspective aligns with the understanding that occupations do not exist in isolation in the economic system, but are instead deeply embedded within and influenced by broader economic cycles, including industry life cycles, technology life cycles as well as many other cycles.
Each of these elements contributes to the evolving nature of occupations, reflecting changes in demand and supply for labor, shifts in relevant skill sets, and the overall economic environment.
The interplay between these cycles can significantly impact the genesis, growth, maturity, and possibly, the eventual decline or transformation of specific occupations.
The OLC thus becomes a concise graphical description, reflecting the compounded effects of these varied cycles, illustrating how shifts in technology, market demands, and demographic changes can precipitate the rise and fall of professional roles and skill sets.

An integrative and interdisciplinary perspective from different levels of economic systems is particularly helpful in understanding the concept of OLC.
In addition, our approach also offers practical insights for policymakers, educators, and individuals navigating the ever-evolving occupational landscape.

\subsubsection{Industry and OLC}
The theory of ILC offers a structured way to understand the evolution of industries through stages of introduction, growth, maturity, and decline~\cite{utterback1975dynamic,klepper1997industry}.
Each stage of the industry life cycle has implications for occupational dynamics within that industry.
In the introduction phase, new occupations may emerge to meet the novel demands of the nascent industry.
As the industry grows, there is an increased demand for labor, leading to the expansion and diversification of occupations.
During maturity, the demand for occupations stabilizes, and specialization within occupations becomes more pronounced.
Finally, in the decline phase, certain occupations may contract or become obsolete as the industry retracts or transforms in response to external pressures.
Thus, the trajectory of the ILC directly influences the evolution of the OLC, necessitating adaptive workforce skills and occupational strategies to navigate these transitional phases.
For instance, smart manufacturing~\cite{ghobakhloo2020industry}, 6G communication~\cite{dang2020should}, and generative AI~\cite{eloundou2023gpts} are among those promising areas of future industries in which new related occupations are still in their infancy.

\subsubsection{Technology and OLC}
The TLC, exemplified by the Gartner Hype Cycle (GHC)~\cite{linden2003understanding,dedehayir2016hype}, further elucidates the relationship between technological advancements and OLC.
The GHC delineates the stages of a technology's life cycle: from the Technology Trigger through Peak of Inflated Expectations, the Trough of Disillusionment, the Slope of Enlightenment, and finally, the Plateau of Productivity.
Each stage signals varying demands for occupational roles, from research and development during the early stages to implementation and maintenance in later stages.
For example, the advent of blockchain technology~\cite{calvao2019crypto,chunmian2022investigating} created a surge in demand for blockchain developers and consultants during its peak of expectation, a demand which later evolved as the technology matured and found realistic applications.
This evolution reflects the direct impact of technological maturation on the demand for specific occupations, highlighting the TLC's pivotal role in shaping the OLC.

\subsubsection{Other cycles and OLC}
Besides ILC and TLC, which are among the most closely related life cycle phenomena regarding OLC, exploring the influence of macroeconomic and human life cycles on the OLC reveals additional layers of complexity.

The macroeconomic cycle, encompassing fluctuations in economic activity such as expansions and recessions, also plays a pivotal role in shaping the OLC through occupational mobility~\cite{moscarini2008occupational}.
During economic upturns, new occupations may emerge as demand for novel roles increases, while downturns often necessitate occupational flexibility and adaption.

Moreover, the human life cycle adds both an individual and aggregate dimension to the OLC narrative.
As individual progress through different stages of their life, their occupational choices and trajectories are influenced by a combination of personal development, family responsibilities, societal expectations, and economic conditions~\cite{super1980life}.
The intergenerational transmission of knowledge and skills, alongside shifts in educational and  career opportunities, reflects the broader socio-economic context in which human life cycle and OLC intersect~\cite{long2013intergenerational,jarvis2017rising}.
The aggregate aging workforce, education, and career progression, further intersects with the OLC by influencing labor market dynamics and occupational transitions~\cite{autor2009job}.

The macroeconomic modelling of overlapping generations provide valuable insights into understanding the concept of OLC~\cite{fougere2007sectoral}.
First, it illustrates how the human life cycle intersect and influence the OLC at both the individual and aggregate levels.
Further, when we treat an occupation as an organic and living object, the overlapping generations model originated from the economics literature~\cite{samuelson1958exact} provides an intuitive description on the process during which successive novel occupations emerge from, interact with, and substitute for the obsolete ones.

\subsection{An economic analysis of the occupational development}
The economic analysis of occupational development from a dual market perspective underscores the intricate relationship between the demand for and supply of occupational talent.
By considering both sides, policymakers, educators, and businesses can better comprehend the forces shaping occupational landscapes and devise strategies to address mismatches between supply and demand.


\subsubsection{Demand factors affecting the occupational development}
The demand for occupational talent is a multifaceted phenomenon, shaped by various economic and societal forces.
Each factor plays a crucial role in determining the kinds of skills and professions that are in high demand at any given time.

First, market demand emerges as a fundamental driver~\cite{murphy1993occupational,rifandi2019stem}, where the overall health of the economy dictates the need for various occupations.
For example, during periods of an economic expansion, there is an increased demand for construction workers to build new infrastructure, financial analysts to manage investments, and sales professionals to leverage new market opportunities.
Conversely, economic downturns may reduce the demand in these areas but increase it for debt counselors or restructuring experts.

Public and societal challenges significantly impact demand at the occupation level~\cite{del2020supply,forsythe2020labor}. This is vividly illustrated in the increased need for healthcare workers, including nurses and doctors, amidst global health crises or the surge in demand for environmental scientists and sustainability experts as climate change becomes a more pressing concern.
These trends reflect society's evolving priorities and the critical role of occupations in addressing these challenges.

Demand driven by technology innovations marks another pivotal area of occupational development~\cite{autor2003skill,webb2019ai,acemoglu2019automation,brynjolfsson2018machinelearning,acemoglu2022artificial}.
The advent of the internet, for instance, created a whole new sector demanding web developers, digital marketers, and cybersecurity experts.
Similarly, the advancements in artificial intelligence and machine learning have spawned demand for AI researchers and practitioners who can navigate these cutting-edge technologies.

Specialization and competition necessitate a deeper dive into the ways business seek to distinguish themselves~\cite{weiss1971learning,law2005specialization}.
In the highly competitive tech industry, for example, companies compete for top talent in software development and data analysis to innovate and outpace competitors.
This drive for specialization not only pushes the boundaries of existing fields but also creates entirely new niches and occupations, such as social media managers or user experience designers.

\subsubsection{Supply factors affecting the occupational development}
On the flip side, the supply of talent within various occupations is influenced by a distinct set of factors, each encouraging or discouraging individuals from pursuing certain career paths~\cite{blau1956occupational}.

Economic reasons play a significant role in guiding career choices~\cite{doepke2008occupational,ghatak2001occupational,banerjee1993occupational}.
Occupations with higher earning potentials as well as risks, such as law, medicine, or engineering, tend to attract many individuals, given the clear economic incentives.
This economic motivation underscores the importance of salary and job stability in shaping the occupational landscape.

Pursuit of passion or prestige illustrate another dimension in occupational and career choice~\cite{amarnani2020consumed,budjanovcanin2022regretting}.
Fields such as academia, arts, and sports attract individuals driven by a deep interest in the subject matter or the honor and recognition associated with excellence in these fields.
For instance, the prestige of being a renowned scientist or the intrinsic satisfaction of creating art motivates individuals to enter and persist in these careers despite potential economic or practical challenges.


Herding effects~\cite{kubler2003information,spyrou2013herding}
reveal the social influences on occupational choices through the information channel.
Trends in career popularity, often amplified by media or peer influence, can lead to a surge in interest towards certain occupations.
The technology boom, for example, has made careers in software engineering or technology entrepreneurship highly desirable, often leading to oversaturation in these fields and undersupply in others, reflecting the powerful impact of societal trends and perceptions on career decisions.

\subsubsection{Integrating supply and demand in OLC analysis}
A comprehensive understanding of the OLC requires a holistic view that incorporates both the supply and demand sides of the labor market at the occupation level.
This dual perspective is essential because the dynamics of occupational development and decline are influenced not only by the evolving needs of the economy and society but also by the changing aspirations, skills, and values of the workforce.
An imbalance between supply and demand can lead to skills shortages, underemployment, or oversupply in certain occupations, negatively affecting economic efficiency as well as individual career satisfaction.

Moreover, behavior-based metrics from online recruitment platforms present a promising avenue for developing nuanced OLC metrics.
These platforms offer a rich dataset encompassing job postings, applicant profiles, hiring trends, and salary information, providing real-time insights into both demand and supply dynamics.
Such metrics are inherently dynamic and reflective of the actual movements within the labor market, making them ideal for capturing the multifaceted nature of the OLC.
By analyzing this data, stakeholders can gain a comprehensive understanding of labor market dynamics, facilitating more informed decisions regarding occupational development and policy-making.

\section{Methodology}\label{sec3}

The OLC model provides a comprehensive framework for tracing the evolution of occupations, from their emergence to eventual decline, as indicated by fluctuations in employment figures within those fields. This strategic approach enables proactive career development and resource allocation, enhancing adaptability in a dynamically changing job market.

\subsection{Definitions of OLC Curve}\label{subsec1}

The OLC curve outlines the evolution of an occupation across five distinct stages: growth, peak, fluctuation, maturity, and decline. These stages are conceptualized based on a combination of empirical data analysis and theoretical frameworks from both the industry life cycle and technology adoption cycles. The accompanying diagram visually summarizes these stages.

\begin{figure}[h]
\centering
\includegraphics[width=0.9\textwidth]{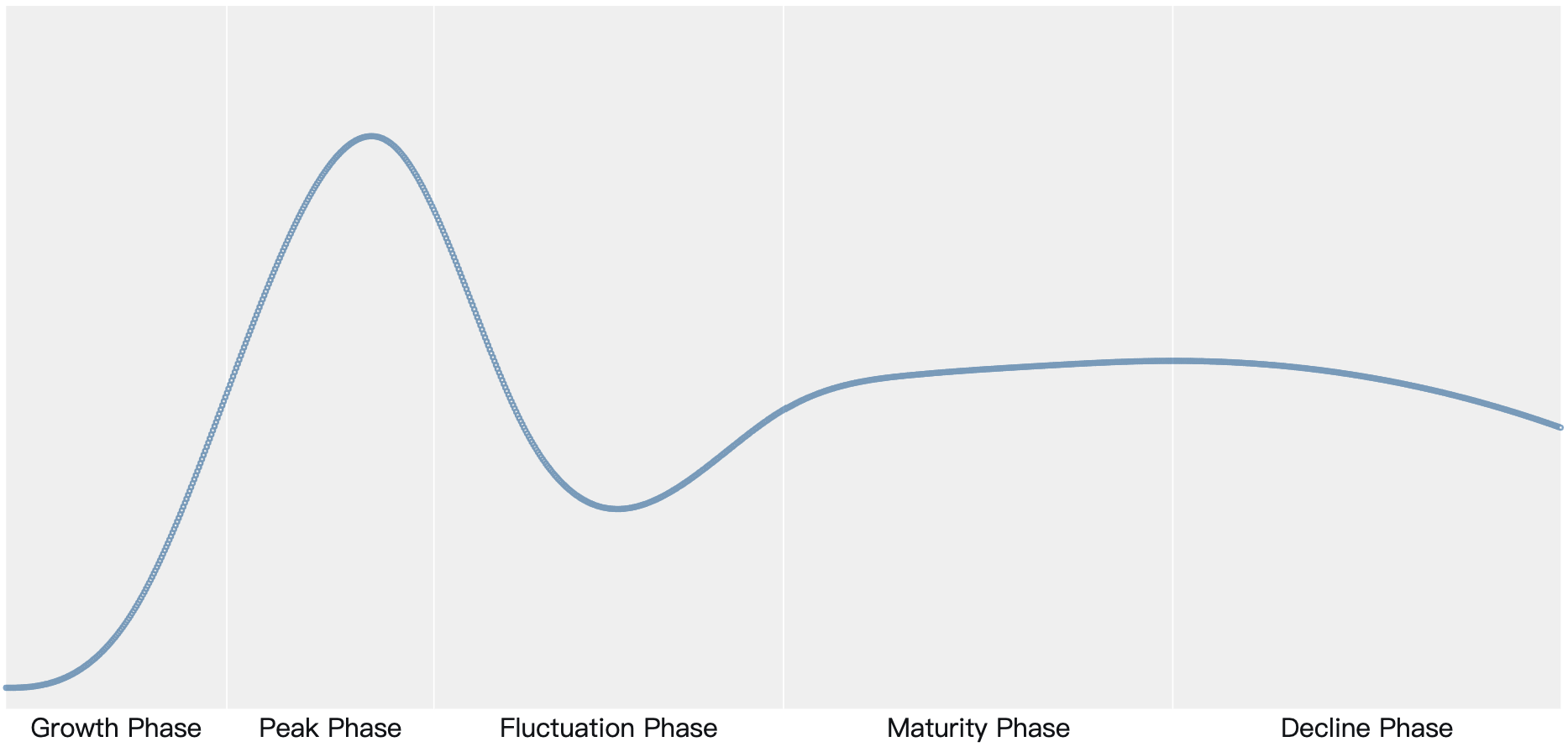}
\caption{Five stages of OLC curve.}\label{fig1}
\end{figure}

\begin{definition}
Growth Phase. The emergence stage of an occupation.
\end{definition}
This initial stage marks the emergence of an occupation, characterized by a notable surge in job demand. The demand for new occupations outstrips the supply of skilled workers, as mastering the required skills takes time. Consequently, salaries typically rise sharply, signaling a period abundant with opportunities yet demanding quick adaptation and learning from jobseekers.

\begin{definition}
Peak Phase. An occupation reaches its highest level of popularity.
\end{definition}
Here, the demand for the occupation reaches its peak, with salaries being high and stable. This period is particularly prosperous for those with high skill levels and experience, though the increase in interest leads to heightened competition.

\begin{definition}
Fluctuation Phase. A period of varying adjustment for an occupation.\end{definition}
At this juncture, demand begins to vary, leading to a stabilization or slight decline in salaries as the market adjusts to the balance between supply and demand. This phase represents a time of transition, encouraging individuals to contemplate their career futures.

\begin{definition}
Maturity Phase. \textit{The stage of stable demand and established presence in the market.}\end{definition}
Demand becomes steady during this phase, mainly driven by industry turnover and the filling of vacancies. The supply-demand equilibrium and salary levels find a balance, though differences in skills and experience can still affect demand for certain roles.

\begin{definition}
Decline Phase. The phase where an occupation faces decreasing demand and relevance.\end{definition}
Triggered by technological advancements or shifts in market demand, this phase sees a reduction in the need for the occupation, with a significant drop in demand and salaries. Oversupply becomes prevalent, urging professionals to consider changing careers or updating their skills.

It is crucial to recognize that not every occupation strictly follows these five stages. Some may undergo multiple peaks due to technological innovations or changes in market demand. For example, tech sector occupations could experience renewed interest with new technological developments like blockchain, artificial intelligence, and the Internet of Things. Conversely, some roles may quickly move from growth to decline without going through substantial fluctuation or maturity phases, as seen in certain manufacturing jobs affected by rapid changes in technology or market dynamics. Despite these variations, OLC serves as a comprehensive tool for grasping the dynamic nature of career development across different fields.

\subsection{Data and constructing process}\label{subsec2}

To investigate the dynamic landscape of occupational evolution, we leverage job posting data from one of the largest recruitment platforms in China. This approach is predicated on the premise that the volume of job postings for a particular occupation serves as a reliable proxy for the demand for labor within that occupation. Numerous studies have supported the use of online job postings as indicators of labor market demand, reflecting both the current workforce needs and future employment trends~\cite{vstefanik2023using}. Given the recent proliferation of online recruitment, which has significantly transformed the job market landscape, our analysis is confined to a relatively short yet insightful period from 2018 to 2023.

Our methodology for determining the current stage of an occupation within its life cycle involves a series of meticulously designed steps. Initially, we extract relevant keywords from the titles of job postings to accurately categorize each occupation. Subsequently, we aggregate the total number of job postings for each occupation annually. This yearly aggregation is chosen to mitigate the distortion caused by seasonal hiring fluctuations, which could obscure the underlying occupational trends.

Following data aggregation, we employ a trend analysis to assign each occupation to a specific stage in its life cycle. This classification is based on the pattern of change in the number of job postings: a continuous increase suggests the growth phase, whereas other patterns correspond to different phases (peak, fluctuation, maturity, decline) of OLC. To automate this classification, we develop and iteratively refine a set of rules that accurately reflect the observed realities of occupational evolution.

In the final step of our analysis, we plot the occupations on a graph, positioning them according to their rate of growth within the relevant phase of the OLC. Occupations with faster growth rates are depicted further back in convex phases and nearer the front in concave phases. It is important to underscore that, within any given stage, the absolute positioning of an occupation lacks intrinsic meaning; rather, it is the relative positioning that illuminates the comparative growth velocity against other occupations.

This analytical framework, underpinned by a robust dataset and a coherent methodological approach, enables a nuanced understanding of occupational trends. By dissecting the stages of occupational evolution and their relative growth speeds, our study contributes valuable insights into the shifting dynamics of the labor market.

\section{Results}\label{sec4}

\subsection{OLC curve of different occupations}\label{subsec1}

Certain occupations exhibit a remarkable level of resilience, such as teachers, doctors, and police officers, which fulfill essential human needs and possess life cycles that may extend over hundreds or even thousands of years. Nonetheless, in today's rapidly evolving environment, we also see the quick progression through the life cycles of some occupations, with certain roles lasting only a decade or merely a few years.

\begin{figure}[h]
\centering
\includegraphics[width=0.9\textwidth]{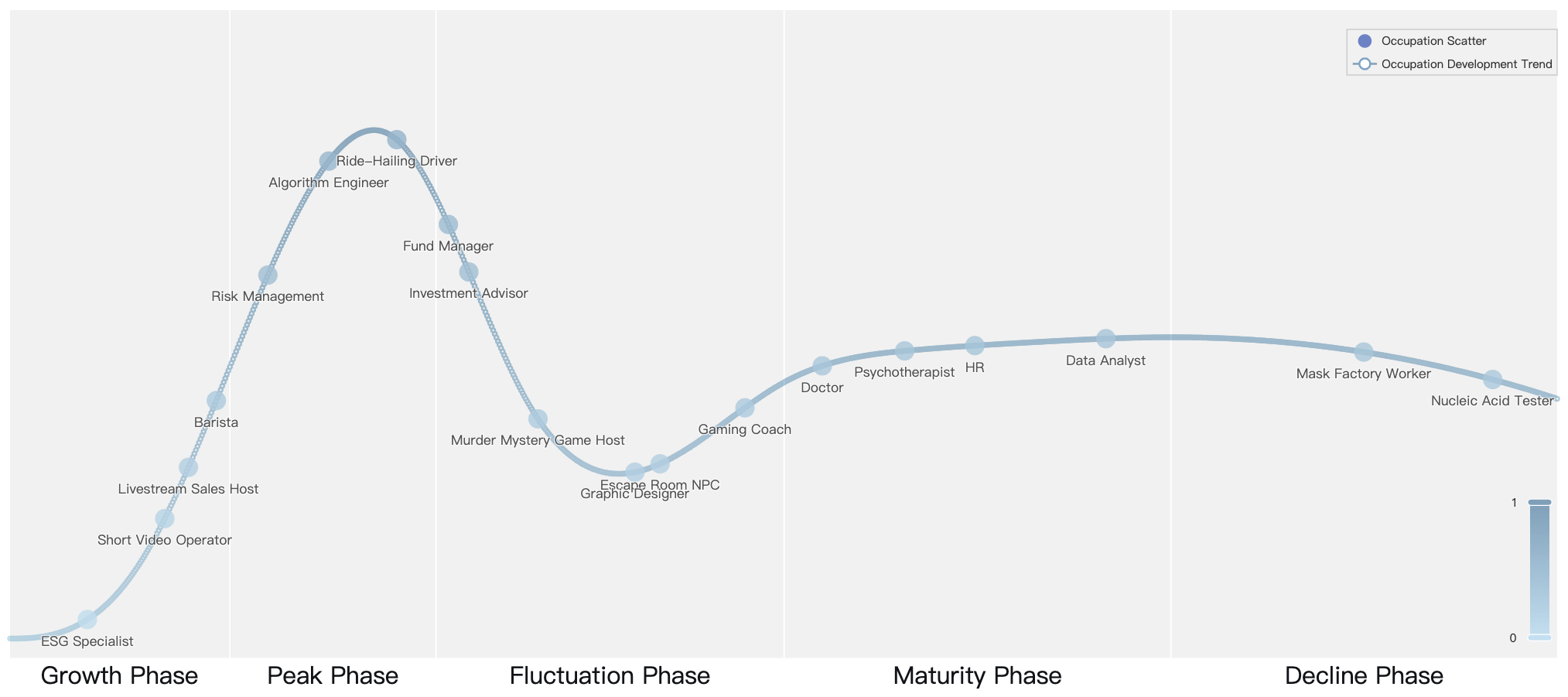}
\caption{OLC curve of different occupations.}\label{fig2}
\end{figure}

To clearly illustrate the concept, we have chosen a selection of typical professions to analyze their positions within the OLC curve, assessing shifts in job demand over recent years. Our analysis reveals that professions such as ESG specialists are in a growth phase, reflecting a national focus on sustainable development. Similarly, the explosion of short video and e-commerce platforms has nearly doubled the demand for roles like short video operators and live stream sales hosts, also marking them as growth-stage occupations.

Conversely, fields like risk management, algorithm engineering, and ride-hailing are experiencing a slowdown, indicating they have reached a peak. At this stage, professions are generally well-recognized and accepted in the market, experiencing stable or mild growth in demand yet contending with saturation and intensifying competition.

In contrast, roles such as script murder game hosts and escape room NPCs, which surged in popularity in 2021, are now witnessing a gradual decline, placing them in a volatile phase. This fluctuation underscores shifting market demands for this entertainment form. Established professions, however, maintain a consistent pattern; doctors, psychologists, HR professionals, and data analysts have seen little change in job demand, akin to the evergreens of the job market, consistently delivering services year after year. Moreover, certain occupations born from specific events phase out as those events conclude, with mask factory workers and nucleic acid testers serving as prime examples.

This analysis highlights the varied dynamics within the occupational ecosystem, emphasizing the impact of technological progress, market demands, and societal changes on the longevity of occupations. It also underscores the importance for individuals to stay flexible and proactive in their career planning, mindful of the potential for swift shifts in job market trends.

\subsection{OLC curve of AI occupations}\label{subsec2}

The surge in ChatGPT's popularity has drawn significant attention to the AI field and its associated occupations. Our examination specifically targets AI-related careers, focusing on roles that are fundamental to AI development rather than those that simply make use of AI technologies.

\begin{figure}[h]
\centering
\includegraphics[width=0.9\textwidth]{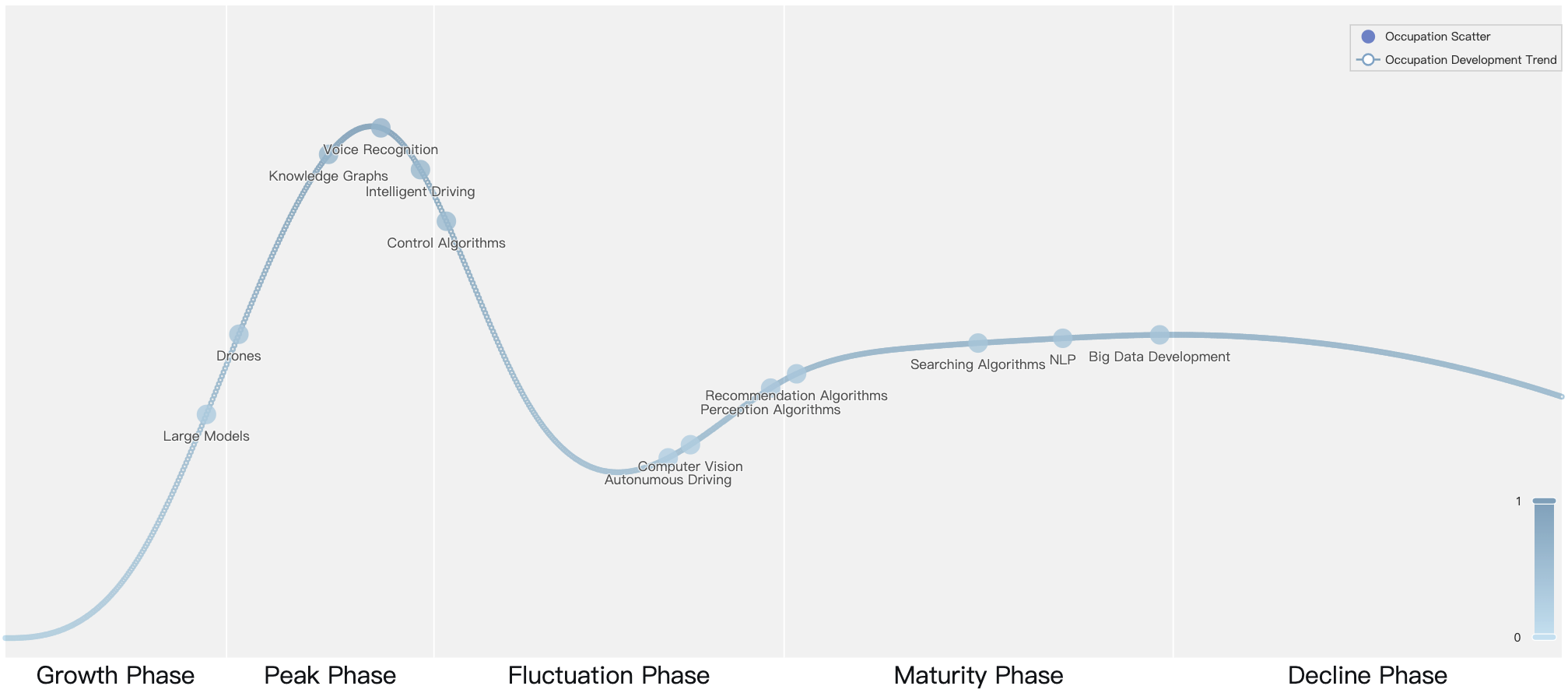}
\caption{OLC curve of AI occupations.}\label{fig1}
\end{figure}

The widespread application of large models in areas such as natural language processing, image recognition, and beyond, coupled with their contributions to enhancing automation and intelligent decision-making, has led to a dramatic increase in demand for professionals in these fields. This surge of interest and investment indicates that AI-related occupations are currently experiencing a growth phase.

Certain technologies, including drones, knowledge graphs, voice recognition, and autonomous driving, are at the peak of their demand. These technologies have transitioned from initial experimentation to broad application in both business and daily life. For example, drones are utilized for agricultural surveillance and package delivery, knowledge graphs improve search engine and recommendation system accuracy, voice recognition technology is integrated into smart assistants and customer support, and autonomous driving is being developed to enhance road safety and efficiency. The successful expansion of these technologies has propelled the demand for professionals in these areas to its zenith.

On the other hand, control algorithms, autonomous driving, machine vision, and perception algorithms are navigating the fluctuation phase. These areas face challenges and uncertainties, including technological bottlenecks and market saturation, necessitating innovation and new application scenarios to foster growth. Moreover, as technological standards evolve, the demand for professionals in these fields experiences variability.

Technologies such as recommendation systems, search algorithms, natural language processing, and big data analytics have reached the maturity phase. These foundational components of the modern digital economy and information society are widely applied across various sectors, including e-commerce, social media, and corporate information systems. As these technologies' development and application stabilize, the demand for and salaries of professionals in these areas also reach a state of equilibrium, showcasing the emergence of mature industrial ecosystems and talent markets.

Considering the dynamic and flourishing state of artificial intelligence as a whole, no AI-related occupations are observed to be in decline. Despite potential fluctuations and challenges, the AI sector remains ripe with opportunities and prospects for innovation and growth.

\section{Conclusion}\label{sec5}

The evolution of occupations in the face of technological innovation and market transformation presents a complex mosaic of opportunities and challenges. Our investigation, anchored in the Occupation Life Cycle (OLC) model, reveals the nuanced pathways that different occupations navigate through their lifecycle stages. The resilience of traditional roles contrasts with the rapid emergence and evolution of occupations in the digital and AI domains, reflecting broader economic and societal trends. In a world of constant change, a thorough grasp of OLC equips us to better anticipate the future, whether it is capitalizing on emerging job sectors or addressing shifts in established ones.

This perspective advocates for a dynamic understanding of occupation evolution, highlighting how technological innovations, economic fluctuations, and societal transformations can quickly change the demand for certain skills and occupations. It underlines the significance of continual learning, adaptability, and the willingness to adapt as necessary. For individuals, this may involve keeping abreast of industry developments, committing to ongoing education, and being receptive to acquiring new skills or enhancing existing ones. For organizations, it stresses the importance of adopting talent management practices that are agile enough to pre-empt changes in skill requirements and promote staff development.

Policymakers also have a crucial role in creating an environment conducive to innovation, offering accessible education and training opportunities, and maintaining a labor market that is agile and inclusive. Such initiatives can ease the transition for individuals moving through the occupational landscape, contributing to a workforce that is both resilient and flexible.

In essence, the OLC is more than just a framework for understanding—it is an impetus for active engagement with the evolving landscape of the professions. By embracing the complexities of career paths, we can collectively strive towards a labor market that not only adapts to change but also flourishes because of it.

\bibliography{sn-bibliography}

\end{document}